\documentclass{article}
\usepackage{spconf,amsmath,graphicx}
\usepackage{amssymb}
\usepackage{pifont}
\usepackage{spconf,amsmath,graphicx}
\usepackage{booktabs}
\usepackage{multirow}
\usepackage{array}
\usepackage{lipsum}
\usepackage{float}
\usepackage{url}
\usepackage{makecell}
\usepackage{hyperref}
\usepackage[caption=false]{subfig}
\usepackage[dvips]{epsfig}
\hypersetup{
     colorlinks=true,
     citecolor=blue,
}

\title{INTERMEDIATE FINE-TUNING USING IMPERFECT SYNTHETIC SPEECH FOR IMPROVING ELECTROLARYNGEAL SPEECH RECOGNITION}
\name{Lester Phillip Violeta, Ding Ma, Wen-Chin Huang, Tomoki Toda}
\address{Nagoya University, Japan}
\begin{document}
\ninept
\maketitle
\begin{abstract}
Research on automatic speech recognition (ASR) systems for electrolaryngeal speakers has been relatively unexplored due to small datasets. When training data is lacking in ASR, a large-scale pretraining and fine tuning framework is often sufficient to achieve high recognition rates; however, in electrolaryngeal speech, the domain shift between the pretraining and fine-tuning data is too large to overcome, limiting the maximum improvement of recognition rates. To resolve this, we propose an intermediate fine-tuning step that uses imperfect synthetic speech to close the domain shift gap between the pretraining and target data. Despite the imperfect synthetic data, we show the effectiveness of this on electrolaryngeal speech datasets, with improvements of 6.1\% over the baseline that did not use imperfect synthetic speech. Results show how the intermediate fine-tuning stage focuses on learning the high-level inherent features of the imperfect synthetic data rather than the low-level features such as intelligibility.
\end{abstract}
\section{Introduction}
\label{sec:intro}

Automatic speech recognition (ASR) is the task of converting speech into its corresponding text transcripts. In the past decade, methodologies that use neural networks have had the most success in this domain \cite{e2e_asr1}. However, while ASR products have been made as commercially available devices, pathological speakers such as those with electrolaryngeal (EL) speech, have not exactly found these devices useful due to their varying speech features \cite{inclusive_asr}. EL speakers suffer from a disrupted larynx, the organ responsible for generating speech, and have it surgically replaced by a mechanical device to recreate the human voice box functions. Electrolarynxes greatly improve the speech intelligibility of the speakers, however, these devices are limited to generating robotic-like speech, which hinders communication \cite{electrolarynx-desc}. Conducting research in developing personalized ASR systems for EL speakers can give them access to the benefits and conveniences of ASR. Unfortunately, only having access to small datasets, coupled with the huge data requirements to train neural networks, has made this a difficult task to resolve.

Common approaches in resolving pathological ASR is through large-scale pretraining with healthy data and subsequently fine-tuning on the target pathological data \cite{pretrain-dysarthric-shor, pretrain-dysarthric-outperforming, pretrain-pathological-lester}. However, in cases where the target data is too small, the domain shift gap between the healthy speech and target data becomes too difficult to overcome, which results in the model being limited to only improving within a certain range. Other typical approaches in resolving minimally-resourced ASR consist of speech synthesis to generate a larger training dataset \cite{gan-adversarial, dysarthric-vocabulary}; however, this approach faces a chicken-and-egg problem, as developing a high-quality speech synthesis system that captures both the acoustic and linguistic information for simulating EL speech would also be very difficult to do without a large dataset. Since the aforementioned literature has revolved around the idea that synthetic data must be as close to the original speech for it to become viable as training data, low-quality synthetic speech has been seen as an unviable dataset alternative. 

However, we hypothesize that an imperfect, low-intelligibility speech synthesis model could instead be used to generate speech that could help close the domain shift gap in the healthy and EL speech when using a large-scale pretraining and target fine-tuning training framework. Common speech synthesis techniques such as text-to-speech or voice conversion use natural inputs like text or speech, making the resulting synthesized speech also inherit the natural, inherent structures of the inputs.
While these inherent and high-level structures in the resulting speech may not be recognizable to the human ear, we hypothesize that a neural network could find these and learn speech representations from the imperfect and distorted synthesized speech, similar to how it would learn when using real speech.

We propose that EL speech recognition performance in neural networks can be improved by simply adding an intermediate fine-tuning step that uses the imperfect synthesized speech after conducting large-scale pretraining. The intermediate fine-tuning should make the network learn high-level acoustic features (i.e., voicing characteristics of EL speech), making it possible for the network to learn the low-level acoustic features (i.e., intelligibility) during the target fine-tuning, closing the domain shift gap between the two datasets.
Through our experiments, we show that this step greatly improves recognition performance in EL speech recognition. Our contributions in this paper are as follows:
\begin{itemize}
  \item We propose a simple intermediate fine-tuning step using imperfect synthetic speech that closes the domain gap between the pretraining and target data and show the success of this by improving the baseline CER by 6.1\%.
  \item We present a thorough study on the effects of using artificially generated EL data in the intermediate fine-tuning step by using different speech synthesis models and distorting text labels in improving CER for EL speech recognition.
  \item We provide a new perspective that imperfect and distorted speech datasets can be useful in training neural networks by instead using it to close the domain shift gap in a large-scale pretraining and fine-tuning training framework. Specifically, we observe that during the intermediate fine-tuning stage, the network ignores linguistic and intelligibility-related information to improve the target fine-tuning.
\end{itemize}

\section{Related works}
\label{sec:nlp_pretraining}
Previous research in natural language processing (NLP) show that pretraining on text data that contains no natural language semantics can be successful in learning language representations, as long as there is some sort of inherent structure in the datasets. For example, \cite{nlp-music, nlp-transferrability-chiang, nlp-transferrability-ri} show that pretraining on music, programming code, or even an artificially generated language using a BERT-based self-supervised method can make an attention-based model such as a long short-term memory or a Transformer learn language representations in an upstream task and perform well in downstream tasks that use natural human language. Even though these datasets are not completely decipherable by a human, they still contain some sort of inherent structure that the attention model could find and use for learning language representations. On the other hand, datasets generated by randomly sampling tokens from n-gram distributions provide little to no benefits to the downstream task. Thus, a model trained on an artificial dataset might be able to learn its inherent structure and could show some improvement on a downstream task. 

\section{Proposed Method}
\begin{figure}[ht]
    \includegraphics[width=8cm]{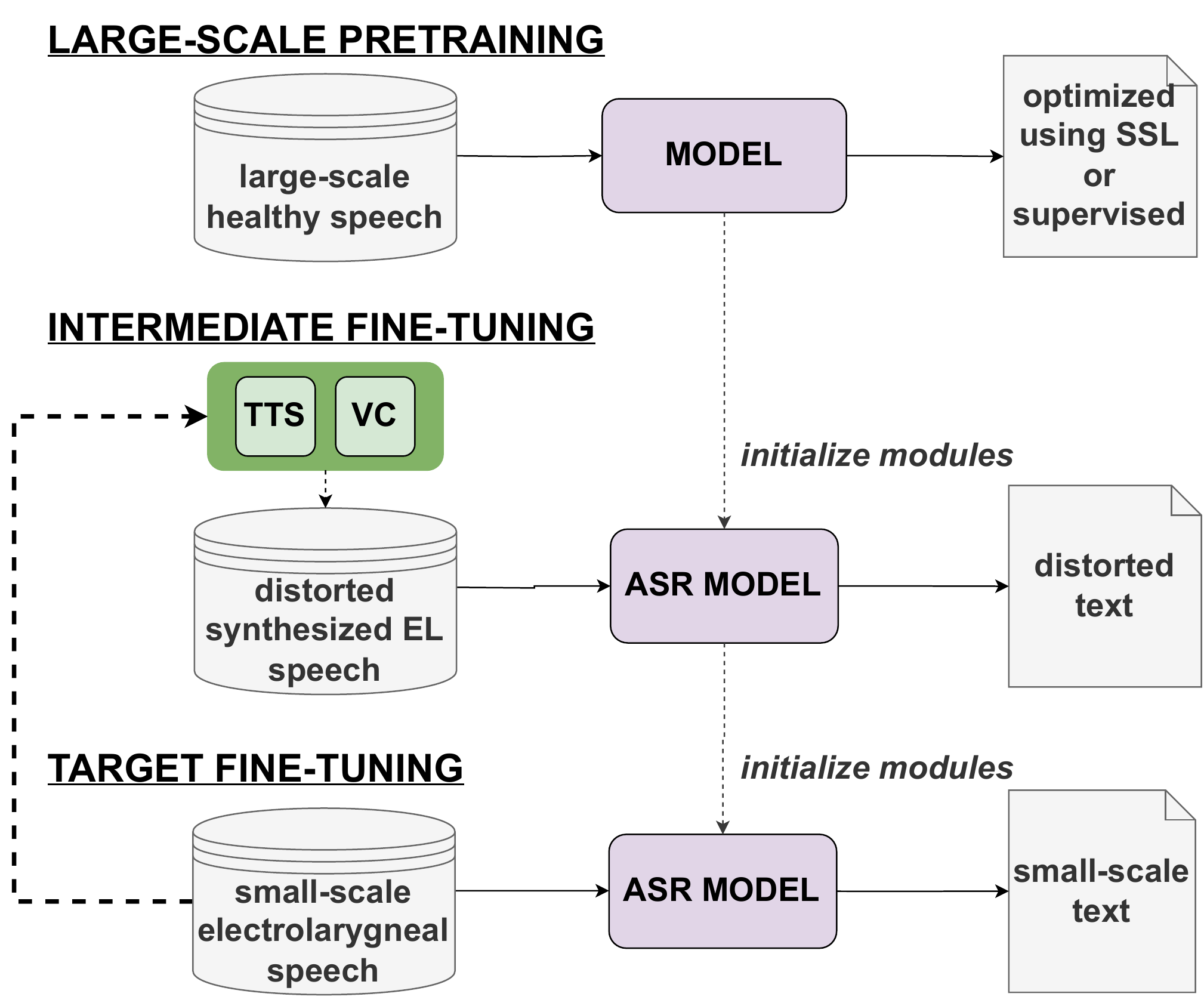}
    \caption{
Visualization of the training steps done in the experiment. We propose that adding in the intermediate fine-tuning stage using imperfectly synthesized speech could help the network improve.
    }
    \label{fig:flowchart}
\end{figure}

Inspired by the research works presented in Section \ref{sec:nlp_pretraining}, we propose that neural networks can also learn EL speech representations from the natural inherent structures in the high-level features found in imperfectly synthesized speech to reduce the domain shift gap. We describe our implementation of the baseline and proposed method below. The overview of the method can be visualized in Figure \ref{fig:flowchart}.

\subsection{Baseline: Large-scale pretraining and target fine-tuning}
We base our baseline method on conventional pathological ASR work such as \cite{pretrain-dysarthric-shor, pretrain-dysarthric-outperforming, pretrain-pathological-lester}, which do large-scale pretraining on healthy speech and fine-tuning on the target EL speech data. The main idea is to use the large-scale dataset to learn general speech representations, and then adapt the model to the features of the target data. We use two types of pretraining methods: supervised and self-supervised. The first pretraining method is supervised pretraining. The model is first pretrained on a large-scale dataset, then fine-tuned on the target data by initializing the weights from the pretrained model. Although the pretraining and fine-tuning data have a mismatch in acoustic features, initializing weights from the pretrained model during the fine-tuning stage allows the model to find the minima relatively faster than starting from random weights, especially if the fine-tuning dataset is not large.

Next, we investigate another pretraining method gaining popularity called self-supervised learning (SSL). An SSL model can be trained without the text labels and using only speech data, which allows it to be trained on a much larger dataset than supervised pretraining. The main advantage in using SSL pretraining is that since we can use a much larger dataset, the model should be able to learn stronger representations that would make it easier for the decoder to convert to text tokens compared to traditional acoustic features such as mel-spectrograms.

\subsection{Proposed method: Intermediate fine-tuning with imperfect EL data}
\label{sec:proposed}
We propose an intermediate fine-tuning step with imperfectly synthesized EL speech in order to close the domain shift gap between the pretraining and fine-tuning data. Imperfectly synthesized speech can be generated using common speech synthesis methods like text-to-speech (TTS) and voice conversion (VC). Details on the synthesis process can be found in Section \ref{sec:implementation_details}. During this step, the model is not expected to improve CER performance, but instead, it should be able to learn the inherent high-level features of the EL speech data that could help it to learn better after fine-tuning with the target EL.

To further show that the model only needs to learn the inherent high-level features and that it does not need to learn linguistic information during the intermediate fine-tuning stage, we distort the target text data labels through two methods. First, we do text-randomization, where we create new text labels of the same lengths by randomly sampling characters from the training text data offline using a uniform distribution. Next, we do text-swapping, where we simply interchange the text information between speech utterances. In the text-swapping setup, both the input speech and target text make sense individually, but are not directly correlated with each other.

\section{Experimental setup}
\subsection{Datasets}
To pretrain the models with large-scale healthy datasets, we used the Japanese LaboroTV Speech dataset (2k hours) \cite{laborotv} for the supervised pretraining setup. For SSL pretraining, we used the Japanese CSJ corpus \cite{csj} (661 hours) and an imperfectly synthesized EL version of the entire CSJ corpus. Since the CSJ corpus is relatively small for SSL pretraining, we also used the large-scale English dataset Librilight \cite{librilight} (60k hours) for comparison.

For our target EL speech during the target fine-tuning stage, we used two parallel datasets, one recorded from a real EL patient and one recorded by a healthy speaker using an electrolarynx, which we will refer to as ELREAL and ELSIMU1 respectively. We also use the healthy speaker version of ELSIMU1 for analysis, referred to as HEALTHY. Next, to train the TTS and VC models, we used another dataset recorded by a different healthy speaker using an electrolarynx, which we will refer to as ELSIMU2, which is entirely non-parallel to the rest of the datasets. 
For the intermediate fine-tuning, we used 20k utterances from the CSJ corpus and converted these to EL using the TTS and VC models. The details regarding the number of minutes and utterances of each dataset can be found in Table \ref{tab:datasets}.

\begin{table}[t]
\centering
    \scriptsize
    \caption{Total duration (in mins) and number of utterances for each split used in our experiment.}
    \label{tab:datasets}
\begin{tabular}{ccccccc}
\toprule
& \multicolumn{3}{c}{\textbf{Minutes}} & \multicolumn{3}{c}{\textbf{No. of Utterances}}\\
\cmidrule(lr){2-4}
\cmidrule(lr){5-7}
\textbf{Dataset} & \textbf{Train} & \textbf{Dev} & \textbf{Test} & \textbf{Train} & \textbf{Dev} & \textbf{Test} \\
\midrule
\midrule
HEALTHY & 4.38 & 2.14 & 2.17 & 116 & 40 & 40  \\
ELREAL & 5.77 & 2.96 & 2.99 & 116 & 40 & 40 \\
ELSIMU1 & 6.26 & 2.71 & 2.75 & 116 & 40 & 40 \\
ELSIMU2 & 125.91 & - & - & 1000 & - & - \\
\bottomrule
\end{tabular}
\end{table}

\subsection{Implementation of speech recognition and speech synthesis models}
\label{sec:implementation_details}
We used ESPnet \cite{espnet-2020}, an open-sourced speech processing toolkit, to implement the speech models. For the supervised ASR model, we used a Conformer \cite{conformer} encoder and a Transformer \cite{transformer} decoder. Each stage used an encoder with eight attention heads and 12 layer blocks, a decoder composed of eight attention heads and eight layer blocks, and was trained by CTC-Attention \cite{ctc-attn2}. For our SSL model, we chose a masked-region prediction framework such as HuBERT \cite{hsu2021hubert}, as this SSL loss previously showed to be more effective in EL speech than a contrastive loss framework \cite{pretrain-pathological-lester}.
Specific details about our supervised and SSL pretraining experiment setups are the same as our previous work in \cite{pretrain-pathological-lester}. We used the character error rate (CER) to evaluate the results. Results shown when using intermediate fine-tuning are the mean of three runs initialized from different random seeds.

For the TTS and VC speech synthesis models, both primarily used the Transformer \cite{transformer} network to create synthetic EL speech using TTS \cite{transformer-tts} or VC \cite{vtn}. The TTS model simply fine-tuned a pretrained model on ELSIMU2. For the VC setup, we followed the same many-to-one setup as described in \cite{n2d}. One slight difference in this experiment's setup though is that since our dataset only contains one source healthy speaker, we generated multiple source healthy speakers using a pretrained TTS model.

\subsection{Generating imperfect synthetic EL data}
\label{sec:synthesis}
We evaluated the TTS and VC models by synthesizing a version of ELSIMU2, and calculated the CER using an ASR model trained on all utterances of ELSIMU2. The synthesized speech resulted in imperfect generations with high CER, with TTS-generated at 34.1\% and VC-generated at 71.9\% despite the ASR model being able to recognize the ground truth speech at only 4.4\% CER. However, even with high CER scores and failing to model the linguistic information, one could still immediately figure out that the speaker speaks with an electrolarynx when listening to the samples, as the model could still somehow model the EL voicing characteristics.

\section{Results and Discussion}
\subsection{Using imperfect synthetic speech in SSL pretraining}
We first investigate the effectiveness of using imperfect EL data as the SSL pretraining data. Although our previous work \cite{pretrain-pathological-lester} has already shown that SSL is not as effective for EL speech, previous research has shown the robustness of SSL models with large improvements in WER when pretrained with in-domain data \cite{robust-w2v}. We use the CSJ dataset and convert it into EL speech using the VC model, which we refer to as CSJ-ELVC, and use the resulting generations as the SSL training data. However, as seen in Table \ref{tab:twostage}, all SSL models from Sys. 1 to 3 still fail to adapt to both EL datasets.

\begin{table}[t]
  \centering
    \scriptsize
    \caption{CER of different pretraining and fine-tuning setups on ELREAL and ELSIMU1. Results with intermediate fine-tuning are mean CER of three runs initialized from different random seeds.}
    \label{tab:twostage}
\begin{tabular}{cllcc}
\toprule
\textbf{Sys.}& \textbf{\begin{tabular}[l]{@{}c@{}}Pretraining Method \\ (Dataset)\end{tabular}} & \textbf{\begin{tabular}[l]{@{}c@{}}Intermediate \\ Fine-tuning Dataset \end{tabular}} & \textbf{\begin{tabular}[l]{@{}c@{}}ELREAL \\ CER\%\end{tabular}} & \textbf{\begin{tabular}[l]{@{}c@{}}ELSIMU1 \\ CER\%\end{tabular}}  \\
\midrule
\midrule
1 &\textbf{SSL (Librilight)} & None & 109.7 & 291.9 \\
2 & \textbf{SSL (CSJ)} & None & 98.4 & 209.0 \\
3 & \textbf{SSL (CSJ-ELVC)} & None & 94.7 & 98.7 \\
\midrule
4 & \multirow{2}{*}{\makecell[l]{\textbf{\begin{tabular}[l]{@{}l@{}}Supervised \\ (LaboroTV)\end{tabular}}}} & None & 22.2 & 17.2 \\
5 & & ELSIMU2 & 21.1 & 15.6 \\
\midrule
6 & \multirow{4}{*}{\makecell[l]{\textbf{\begin{tabular}[l]{@{}l@{}}Supervised \\ (LaboroTV)\end{tabular}}}} & \textbf{TTS, text-randomized} & \textbf{20.2} & \textbf{24.4} \\
7 & & \textbf{TTS} & \textbf{18.7} & \textbf{18.1} \\
8 & & \textbf{VC} & \textbf{18.3} & \textbf{18.0} \\
9 & & \textbf{TTS, text-swapped} & \textbf{16.1} & \textbf{15.7} \\

\bottomrule
\end{tabular}
\end{table}
\label{sec:results}

\subsection{Effectiveness of learning from imperfect speech synthesis during intermediate fine-tuning stage}
\label{sec:results_effectiveness}
As seen in Table \ref{tab:twostage}, adding in an intermediate fine-tuning step improves performance of conventional ASR pretraining methods, proving that we can close the gap in the domain shift between the pretraining and target data using imperfect speech during supervised pretraining. We first observe that although the VC-generated data discussed in Section \ref{sec:synthesis} had a significantly higher CER (71.9\%) than the TTS-generated (34.1\%), there is no CER degradation seen between Sys. 7 and Sys. 8. Thus, we show that CER is not an effective proxy in filtering out EL synthetic data for the training dataset, contrary to the technique in \cite{parrotron}, primarily because pathological speech can be both unintelligible while also natural at the same time.

In addition to this, we observe that distorting text labels can be effective. As seen in Sys. 9, using text-swapped labels proves to be the most effective in the intermediate fine-tuning step in both datasets. However, when using randomly sampled text labels as seen in Sys. 6, we see a lesser effectiveness in the method. This is similar to the findings found in the NLP experiments in Section \ref{sec:nlp_pretraining}, where using pretraining text data randomly sampled from n-grams was also not effective in improving the NLP model performance. Thus, there must be other high-level features that the model was using to improve the target fine-tuning step rather than using the low-level features such as intelligibility.

Another observation is the difference in the results with ELSIMU1 and ELREAL. We see that the intermediate fine-tuning was not as effective and that using the conventional pretraining and fine-tuning method in Sys. 4 results in a lower CER (17.2\%) than the other setups. This is most likely due to ELSIMU1 having a smaller domain shift gap with the healthy pretraining dataset, compared to ELREAL. As ELSIMU1 was only recorded by healthy speakers using an electrolarynx, ELSIMU1 most likely mimicked the robotic quality of EL speech well, but not the other aspects such as the speech rate, pronunciation tendencies, and the such that are caused by the removal of the larynx.

\subsection{Network focuses on finding the inherent structures during intermediate fine-tuning stage}
As the goal of the research is to improve ASR for real EL speakers, we focus more on ELREAL in the discussion of the results. One hypothesis why the network works despite having distorted target linguistic labels is that the network instead optimizes to learn the inherent structures within the speech and text itself and does not prioritize learning the speech-to-text alignment as much as it usually would. For example, since the setup with text random target labels were randomly sampled and each character had no correlation with each other, the intermediate fine-tuning was less effective because there was no pattern for the model to learn. On the other hand, the setup with text-swapped labels had a pattern that grammatically correlated the characters with each other, which the network could use to find and learn to improve the performance in the target EL fine-tuning step. We can relate this to the success of using imperfect EL speech, since the imperfect speech data were generated by a sequence-to-sequence model, each frame had an inherent structure that correlated them with each other. In this case, the high-level features may have been able to represent the voicing characteristics well enough, which the network found and helped it to learn better during the target fine-tuning step.

\begin{figure}[ht]
    \includegraphics[width=8.5cm]{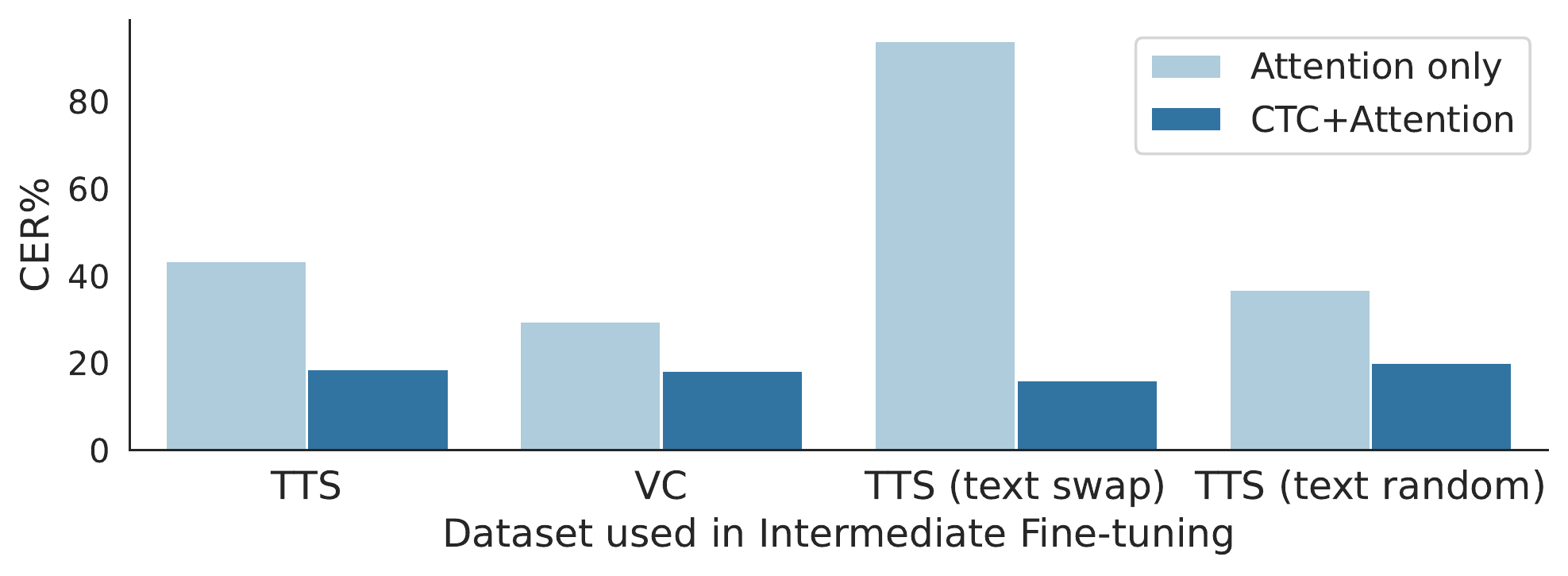}
    \caption{
Visualization of CER difference on ELREAL after the target fine-tuning when (1) using only an attention loss during intermediate fine-tuning, and when (2) using the multi-objective CTC-Attention loss during intermediate fine-tuning.
    }
    \label{fig:loss}
\end{figure}

Although we see success in using text-swapped labels and hypothesize that the intermediate fine-tuning focuses on finding the inherent structures, this is not to say that the network completely disregards the speech-to-text alignment. In Fig. \ref{fig:loss}, we find that if we remove the CTC loss constraint during the intermediate fine-tuning stage, the final CER degrades in all setups. Thus, the network still needs to learn a monotonic alignment using the CTC loss, even with a weak one. This shows that even though there is not a full correlation between the speech and text, learning a weak alignment could still benefit the network. These observations lead us to think that the network may indeed simply be finding correlations in the high-level features, such as the aforementioned inherent structures, rather than the low-level intelligibility features that it usually uses in high-quality datasets.

\begin{figure}[ht]
    \includegraphics[width=8.5cm]{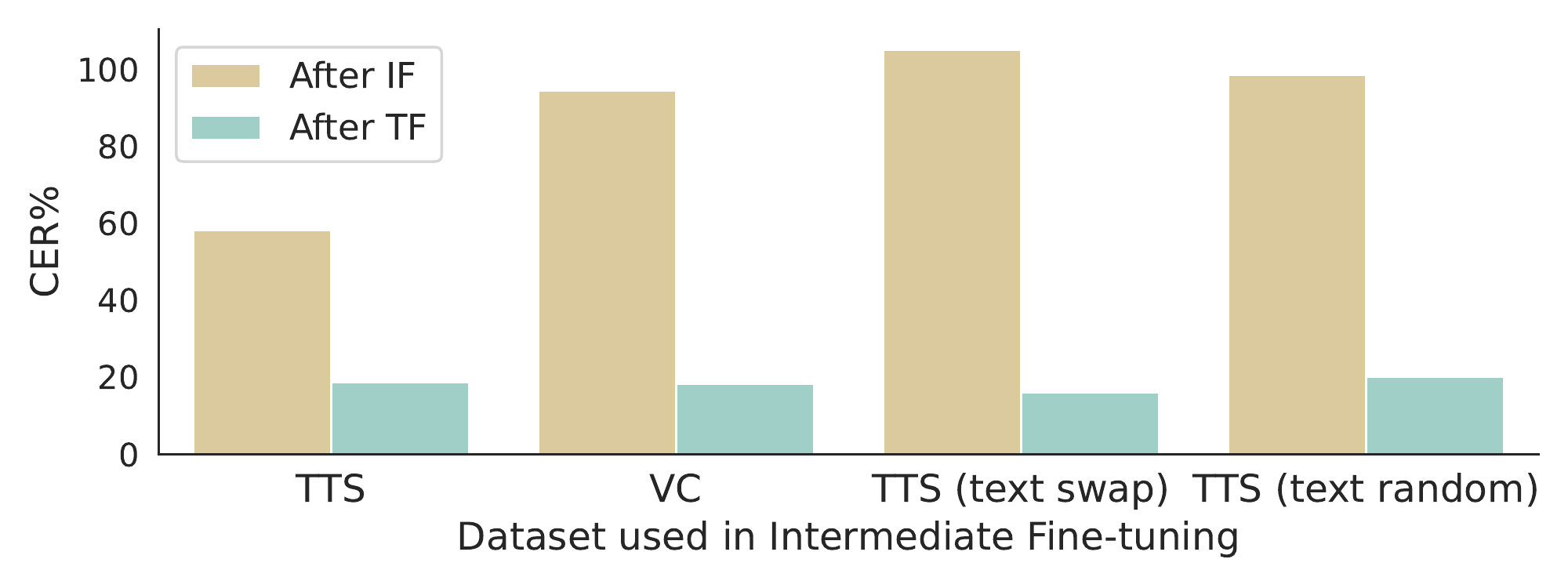}
    \caption{
Visualization of the ELREAL CER  (1) immediately after the intermediate fine-tuning step (IF), and (2) after the target fine-tuning (TF) step.
    }
    \label{fig:cer_steps}
\end{figure}

Building on these observations, another interesting finding in Fig. \ref{fig:cer_steps} that we find is that using the imperfect synthetic speech first results into CER scores even higher than what we get after the large-scale pretraining stage at 77.1\% CER. Despite this initial degradation by the intermediate fine-tuning, the network performance still improves after the target EL fine-tuning step, which proves our initial hypothesis in Section \ref{sec:proposed}. In particular, we see that the text swap, which had the worse CER after the intermediate fine-tuning stage, performed the best after the target fine-tuning, leading us to think that the network indeed does not focus on using low-level intelligibility features during the intermediate fine-tuning stage.

\subsection{Visualizing the latent spaces in each step using t-SNE}
\begin{figure}[ht]
    \includegraphics[width=9cm]{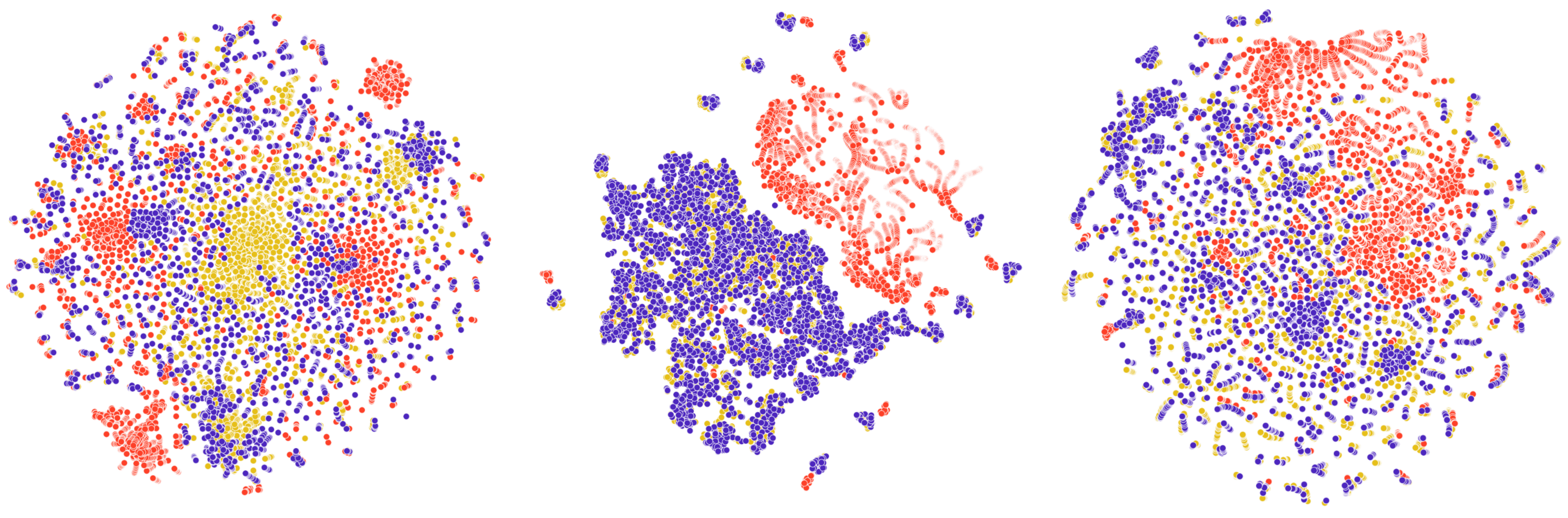}
  \caption{Comparison of how the ELREAL, ELSIMU1, and HEALTHY test sets are projected into the latent space, visualized by t-SNE \cite{t-sne}. Illustrated from left to right are the latent representations after the large-scale pretraining, intermediate fine-tuning (text-swapped), and target data fine-tuning. Each feature from the latent space is represented as a single dot, with orange dots representing healthy speech, violet dots representing ELSIMU1, and yellow dots representing ELREAL.}
  \label{fig:t-sne}
\end{figure}

We visualize the latent representations from each step using t-SNE \cite{t-sne}. To do this, we use the test set of ELREAL, ELSIMU1, and HEALTHY, and extract the latent representation from each utterance. Each frame-level feature is represented as a single dot. As shown in Fig. \ref{fig:t-sne}, we see that intermediate fine-tuning projects each speaker in different spaces, whereas the pretraining and target fine-tuning steps project both speakers in roughly the same space. This shows that the intermediate fine-tuning step optimizes finding the high-level inherent structures by learning to separate real EL speech representations from healthy speech representations, despite never seeing the real EL data before. Moreover, we see that its projections of the data is the opposite of the pretraining and target fine-tuning stages, which focused on improving the CER.

\section{Conclusion}
\label{sec:conclusion}
We showed that adding an intermediate fine-tuning step by using imperfectly synthesized speech, even with distorted text, can be used to close the domain shift gap between pretraining and fine-tuning data. Since the synthetic data inherit the natural structure of their inputs, a neural network could find and use these inherent structures to learn speech representations, even though these said structures would not be apparent to the human ear. Future work could focus on controlling features in the synthetic data that could represent the high-level inherent features such as pitch, timbre, or speaking rate.

\noindent\textbf{Acknowledgements} This work was partly supported by AMED under Grant Number JP21dk0310114, Japan, and a project, JPNP20006, commissioned by NEDO. We would also like to thank Dr. Bence Mark Halpern for his insightful comments in writing this paper.


\bibliographystyle{IEEE}
\bibliography{refs}

\end{document}